\documentclass[fleqn,10pt]{wlscirep}
\usepackage[utf8]{inputenc}
\usepackage[T1]{fontenc}

\usepackage{graphicx}% Include figure files
\usepackage{dcolumn}% Align table columns on decimal point
\usepackage{bm}% bold math
\usepackage[utf8]{inputenc}
\usepackage[T1]{fontenc}
\usepackage{mathptmx}
\usepackage{etoolbox}
\usepackage{amsmath}
\usepackage{subcaption}
\usepackage{textcomp}
\title{Advanced Ceramic Plasma Discharge Capillaries for high repetition rate operation}

\author[1,2,*]{Lucio Crincoli}
\author[1]{Romain Demitra}
\author[1]{Valerio Lollo}
\author[1]{Donato Pellegrini}
\author[3]{Marco Pitti}
\author[1]{Lucilla Pronti}
\author[1]{Martina Romani}
\author[1]{Massimo Ferrario}
\author[1]{Angelo Biagioni}
\affil[1]{INFN - Laboratori Nazionali di Frascati, Frascati (RM), 00044, Italy}
\affil[2]{University of Rome Sapienza, Piazzale Aldo Moro 5, 00185 Rome, Italy}
\affil[3]{University of Palermo, Palermo (PA), 90133, Italy}
\affil[*]{lucio.crincoli@lnf.infn.it}

\begin{abstract}

  In view of future applications of plasma-based particle accelerators, within the fields of high-energy physics and new light sources, the capability of plasma sources to operate at high repetition rates is crucial.
  In particular for gas-filled plasma discharge capillaries, which allow direct control over plasma properties, a key aspect is the longevity of the material, subject to erosion due to the heat flux delivered by high voltage plasma discharges.
  In this regard, we present an innovative design of discharge capillaries based on the use of different ceramic materials, which can sustain high voltage plasma discharges at high repetition rate and, moreover, be easily machined for the complex geometries required for plasma-based accelerators.
  Experimental campaigns are carried out at 10-150 Hz, assessing the longevity of ceramic capillaries by means of different diagnostic techniques.
  In addition, numerical simulations are performed to analyze the heat transfer within the whole plasma source.
  Results from experimental and numerical analysis highlight the capability of ceramic capillaries to preserve plasma properties and the integrity of the source during long-term plasma discharge operation at high repetition rate. In particular, we demonstrated the suitability of the proposed solution for the operative range of 100-400 Hz, foreseen for EuPRAXIA@SPARC\_LAB project.
\end{abstract}
\begin{document}

\flushbottom
\maketitle

%\linenumbers

\section*{Introduction}

Novel particle accelerators based on plasma technology allow a drastic reduction in size and cost, due to the strong GV/m accelerating fields generated in the so-called Plasma Wakefield Acceleration mechanism~\cite{PhysRevLett.43.267}. Among the various plasma sources used for particle acceleration, gas-filled plasma discharge capillaries represent one of the most compact and cost-effective devices for plasma creation and confinement, achieved by means of high voltage pulses~\cite{PhysRevE.63.015401}.
Moreover, plasma discharge capillaries can be used for  particle beam focusing, alternatively to conventional quadrupoles, due to the strong kT/m focusing fields produced during the current discharge~\cite{PhysRevLett.121.174801}.
The compactness of plasma-based accelerators is driving interest in the scientific community concerning a variety of applications, from FEL and light sources~\cite{Assmann} to high energy physics and medical applications~\cite{BOURHIS201918}, however the implementation of such plasma sources in high energy physics experiments and light sources requires long term operation in the high repetition rate regime (from hundreds Hz to over kHz), such as for the EuPRAXIA@SPARC\_LAB project\cite{FERRARIO2018134}, designed to be a plasma-based FEL user facility operating in the range of 100-400 Hz.
In this regard, several issues arise from the operation of the plasma module beyond few Hz, including limitations related to the vacuum pumping system, the high voltage circuit and the plasma source itself.
In particular, concerning gas-filled plasma discharge capillaries, a key point is the capability of the capillary material to dissipate the thermal load produced by high voltage discharges and deposited onto the capillary walls.
%, which is enhanced at higher repetition rate operation.
%Alternative plasma sources can be employed for high repetition rate operation, including gas jets~\cite{Faure_2019} and HOFI plasma channels~\cite{Alejo_2022, PhysRevE.102.053201, PhysRevE.97.053203, PhysRevAccelBeams.22.041302}, which rely on the use of laser pulses for the plasma formation.
%In such sources, the absence of a solid material for the plasma confinement directly removes the capillary overheating issue, but it prevents an easy and controlled confinement of the plasma distribution.
%On the other hand, gas cells~\cite{AUDET2018383, 10.1063/5.0009632} allow a more precise tailoring of the plasma channel, but the damage caused by laser-target interaction prevents high repetition rate operation, similarly to capillary discharges. Moreover, on a general ground, the inclusion of a laser system for the plasma formation adds complexity and expensiveness to the overall plasma module, compared to plasma discharge capillaries.
In order to overcome this issue, enhanced in the high repetition rate regime, high melting temperature materials, such as sapphire, have been proposed and tested~\cite{10.1063/1.4940121}.
On the other hand, the significant production cost and poor machinability of such materials for long and complex geometries can dramatically affect the cost-effectiveness of the source.
In this context, we present the design of plasma discharge capillaries based on ceramic materials, such as Shapal Hi M Soft\texttrademark\cite{Shapal} and Macor\textregistered, characterized by high thermal conductivity, high melting temperature and good machinability. % than sapphire.
Experimental campaigns are carried out from 10 Hz up to 150 Hz on a plasma discharge capillary made in Shapal, inserted in a Macor holder, in order to assess the longevity of such materials exposed to high voltage plasma discharges operating in the high repetition rate regime.
Generally, for capillaries made in transparent materials, such as glass or sapphire, longitudinal diagnostics\cite{PhysRevA.43.5568, Keilmann_1972, Biagioni_2021} are adopted to study plasma properties. However, due to Shapal opacity, alternative diagnostic techniques are implemented to characterize the plasma discharge, such as the transverse Stark broadening method for plasma density measurements. Furthermore, an optical stereomicroscope and a Compact Laser Module (Thorlabs) are employed to monitor the capillary profile during the experimental campaign.
In addition, heat transfer numerical simulations are performed with COMSOL Multiphysics \cite{COMSOL}, to provide theorical support to the experimental analysis.
Results from experimental campaigns prove that, after 20 million discharges at 10-150 Hz, no modification takes place in the whole source and, in addition, plasma properties are well preserved. Moreover, numerical simulations extend the reliability of ceramic capillaries up to 300-400 Hz, beyond which the melting temperature of Macor is reached. Finally, such results perfectly meet the requirements of
EuPRAXIA@SPARC\_LAB project, designed
to operate a plasma-driven FEL source at 100-400 Hz. Experimental and theoretical activities have been carried out at the SPARC\_LAB test facility\cite{FERRARIO2013183}, within the EuPRAXIA framework\cite{Assmann}, and at DAFNE-Light Laboratory\cite{Dafne-luce} at Laboratori Nazionali di Frascati (INFN).

\section*{Methods and materials}

\subsection*{Capillary design}
%\section{Materials and methods}

\begin{figure}[h!]
\centering
\includegraphics[width=0.85\linewidth]{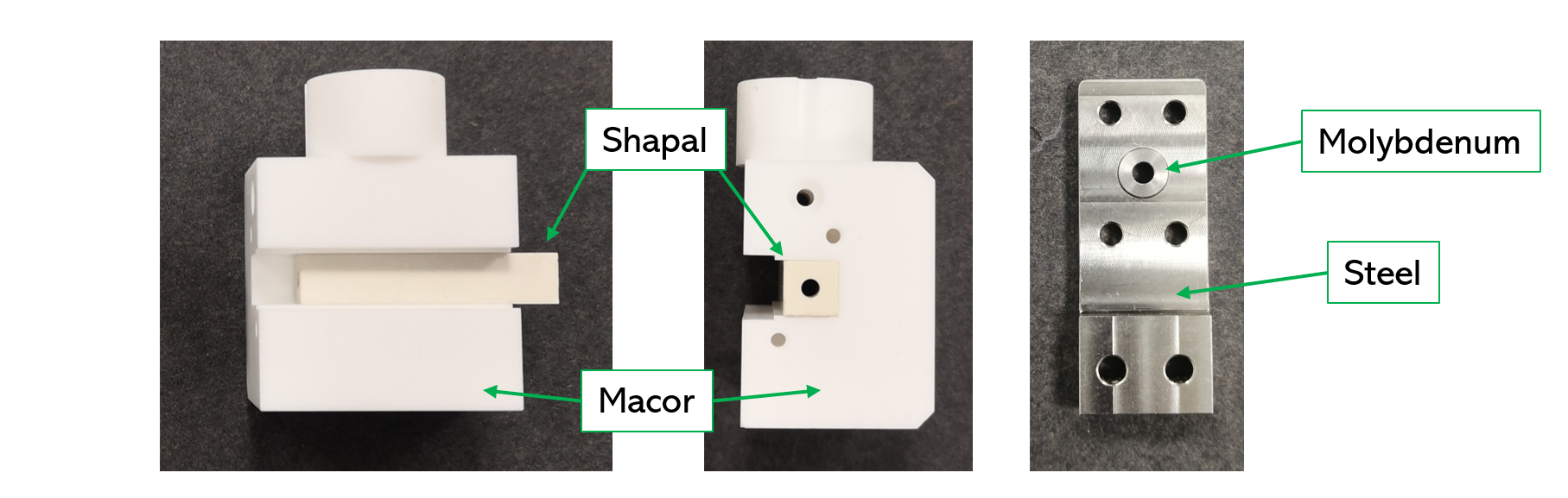}
\caption{Front view and side view of the Shapal capillary, with Macor holder and molybdenum-stainless steel electrodes}
\label{Capillary pictures}
\end{figure}

Plasma discharge tests at high repetition rate are carried out adopting a plasma discharge capillary made in Shapal Hi M Soft, a hybrid type of machinable Aluminum Nitride (AIN) ceramic, mixed with Boron Nitride (BN).
Shapal is characterized by high thermal conductivity (92 W/(m$\cdot$K) at room temperature, 35 W/(m$\cdot$K) at 1000°C), more than two times higher than sapphire, and high melting temperature (1900 °C in vacuum). %, hence it represents a promising material for high-temperature resistant capillaries.%, able to dissipate the heat produced by high repetition rate HV plasma discharges.
Moreover, its good machinability allows the realization of long and thin structures, required for plasma discharge capillaries.
The geometry of the designed Shapal capillary, reported in Fig.~\ref{Capillary pictures}, is characterized by a 3 cm-long channel, with a 2 mm-diameter circular hole, and two inlets with 1 mm-diameter for a uniform gas injection\cite{Crincoli2024CharacterizationOP}.
The capillary is inserted and glued inside a ceramic holder made of Macor, a machinable glass-ceramic having a thermal conductivity of 1.46 W/(m$\cdot$K) at room temperature and 1000°C maximum operating temperature. In particular, Macor is adopted for its excellent machinability, cost-effectiveness and availability for large geometries, allowing the design of long capillary holders that can be easily drilled and machined.
Two electrodes, constituted by an inner molybdenum ring and an outer stainless steel plate, are screwed to the extremities of the Macor holder.
In such configuration, the higher melting temperature metal (2623°C for molybdenum) is directly exposed to the plasma discharge, while stainless steel is used for the outer plate due to its good machinability, exploited for attaching the electrode to the ceramic capillary.

\subsection*{Experimental setup}

\begin{figure*}[h!]
\centering
\includegraphics[width=0.8\linewidth]{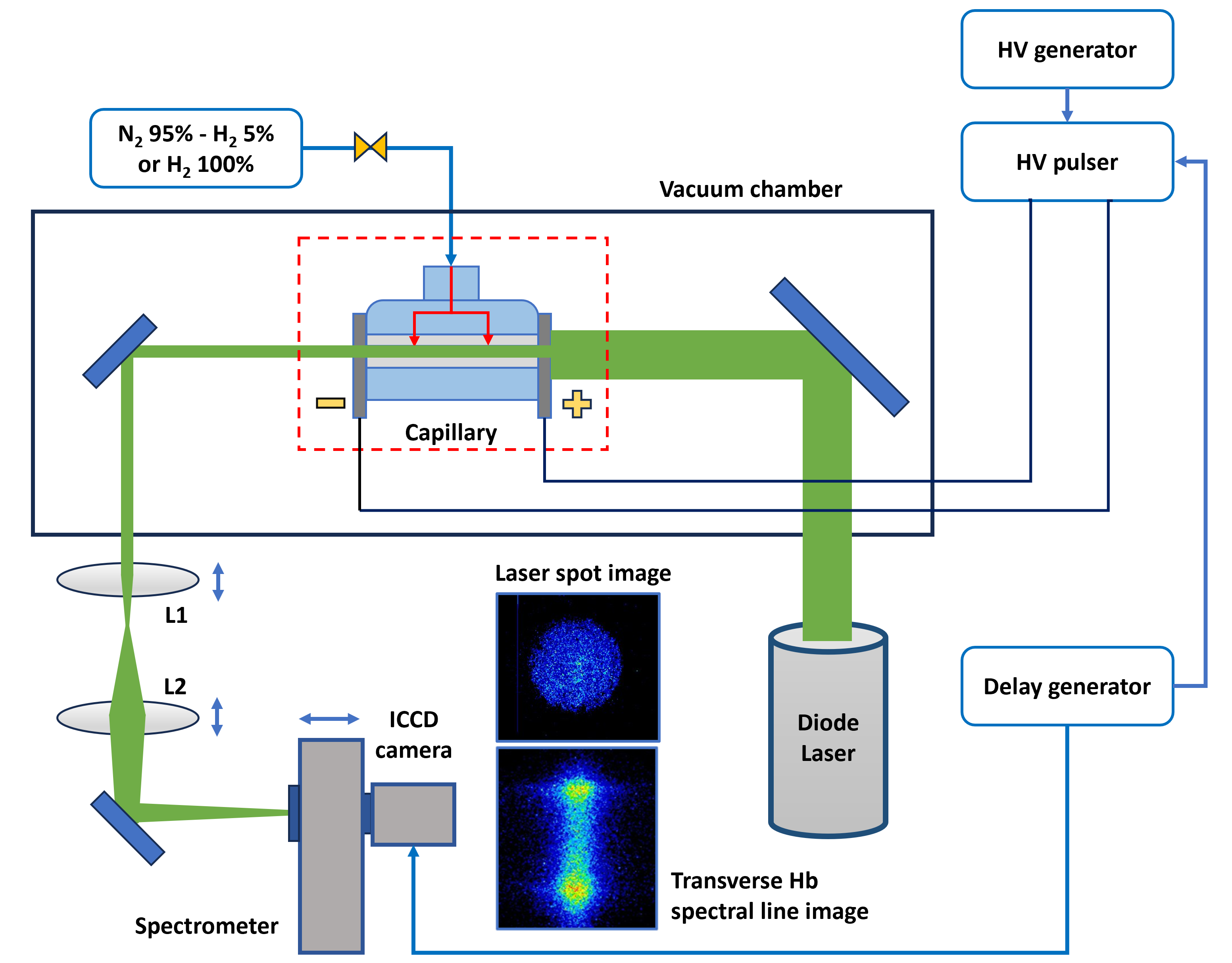}
\caption{Schematic representation of the experimental setup}
\label{Experimental setup}
\end{figure*}

The experimental apparatus, set up for plasma discharge tests at high repetition rate, is constituted by three different systems for the gas injection, plasma formation (HV discharge circuit) and characterization (diagnostics), as reported in Fig.~\ref{Experimental setup}.

The plasma discharge capillary is installed inside a vacuum chamber and filled with a mixture of N$_2$ (95\%) and H$_2$ (5\%) gas, injected in continuous flow regime. %Pulsed gas injection, which would allow better vacuum conditions, is not feasible for the experiment, since the slow activation time of electro-mechanical valves %and the long refilling time of the capillary prevents operation above 10 Hz.
Two primary scroll pumps and one turbo-molecular pump keep a vacuum level of $10^{-2}$ mbar inside the vacuum chamber, suitable to produce and confine plasma discharges.
%without the need to accelerate particle beams.
A high voltage (HV) generator feeds a high voltage electrical pulser, which delivers kV-range $\mu$s pulses to a couple of electrodes, attached to the capillary extremities, in order to ionize the neutral gas mixture inside the capillary channel and produce the plasma discharge~\cite{10HV}.
%In addition, an oscilloscope is used to acquire the waveform of the discharge current.
A delay generator (Stanford Research DG535) sets the operating repetition rate in the range of 10-150 Hz. The upper limit is imposed by the maximum current delivered by the HV generator to charge the pulser circuit.
%the overheating of the HV pulser.
The turbo-molecular pump and the HV pulser are respectively equipped with water cooling and fan cooling systems, allowing high repetition rate operation in thermal steady-state conditions.
As voltage pulses trigger the plasma discharge inside the capillary, the resulting current is measured by means of a Pearson 110 probe, which converts the current pulse into a voltage signal with a transfer impedance of 0.1 V/A.
The voltage signal is then acquired by a 300 MHz WaveAce 2032 oscilloscope.
The waveform of the discharge current pulse is retrieved by applying the 0.1 V/A factor to the voltage pulse acquired by the oscilloscope.

Due to Shapal opacity to visible and near-infrared light, longitudinal plasma density distribution is not measurable by means of conventional spectroscopic or interferometric techniques, therefore transverse diagnostics are employed to characterize the plasma source.
First, as shown in Fig.~\ref{Experimental setup}, a diode laser is employed to monitor any geometrical variation in the capillary channel cross section, eventually caused by Shapal erosion.
The laser beam enters the capillary with a spot size larger than the channel diameter and is guided through a beam transport line, made of a telescopic system (L1 and L2 lenses), into an imaging spectrometer (SpectraPro 275), equipped with an intensified CCD Camera (Andor Istar 320). A mirror implemented inside the spectrometer guides the laser beam into the ICCD camera. In this way, the shape of the laser spot image, acquired by the camera, is directly related to the capillary wall profile.
In addition, the electrodes are characterized by a hole slightly larger than the 2 mm wide capillary diameter, so as to prevent the interference of electrodes erosion in the measurement reliability.
This technique does not allow to detect a channel widening in the inner portion of the capillary, eventually caused by erosion, if the edges do not experience the same deformation, therefore a microscopic analysis is performed by means of an optical stereomicroscope (Euromex) to characterize the entire channel profile with higher precision.

An accurate characterization of the plasma discharge is achieved by means of the spectroscopic analysis of the plasma-emitted light, including the transverse Stark broadening method \cite{Griem2} adopted to measure the plasma density distribution inside the capillary channel.
The same optical line used for laser spot imaging guides the plasma-emitted light from the plasma channel to the imaging spectrometer.
In this way, hydrogen emission lines of the Balmer series, H\textsubscript{$\alpha$} (656.3 nm) and H\textsubscript{$\beta$} (486.1 nm), are selected by the spectrometer diffraction grating (600 grooves/mm) and the spectral line broadening is measured to recover the electron plasma density, exploiting the direct proportionality between the hydrogen spectral line width $\Delta\lambda\textsubscript{1/2}$ and the electron plasma density $N\textsubscript{e}$  , according to the Stark effect:

\begin{equation}\label{Stark}
N\textsubscript{e} = 8.02\times10^{12} \left[ \frac{\Delta\lambda\textsubscript{1/2} [\text{Å}]}{\alpha\textsubscript{1/2} [\text{Å}]} \right]^{3/2} [cm^{-3}]
\end{equation}

in which $\alpha\textsubscript{1/2}$\cite{Griem1} depends on the plasma density, assumed around $10^{17}$\ $cm^{-3}$, and the plasma temperature, estimated by means of the quasi-static model of the plasma discharge, developed by Bobrova\cite{Bobrova}:

\begin{equation}\label{Bobrova}
T\textsubscript{e} = 5.7 \left[ \frac{I[kA]}{r\textsubscript{0} [mm]} \right]^{2/5} [eV]
\end{equation}

with $I$ the discharge current measured by the oscilloscope and $r_0$ the radius of the capillary channel.

\begin{figure*}[h!]
\centering
\includegraphics[width=0.55\linewidth]{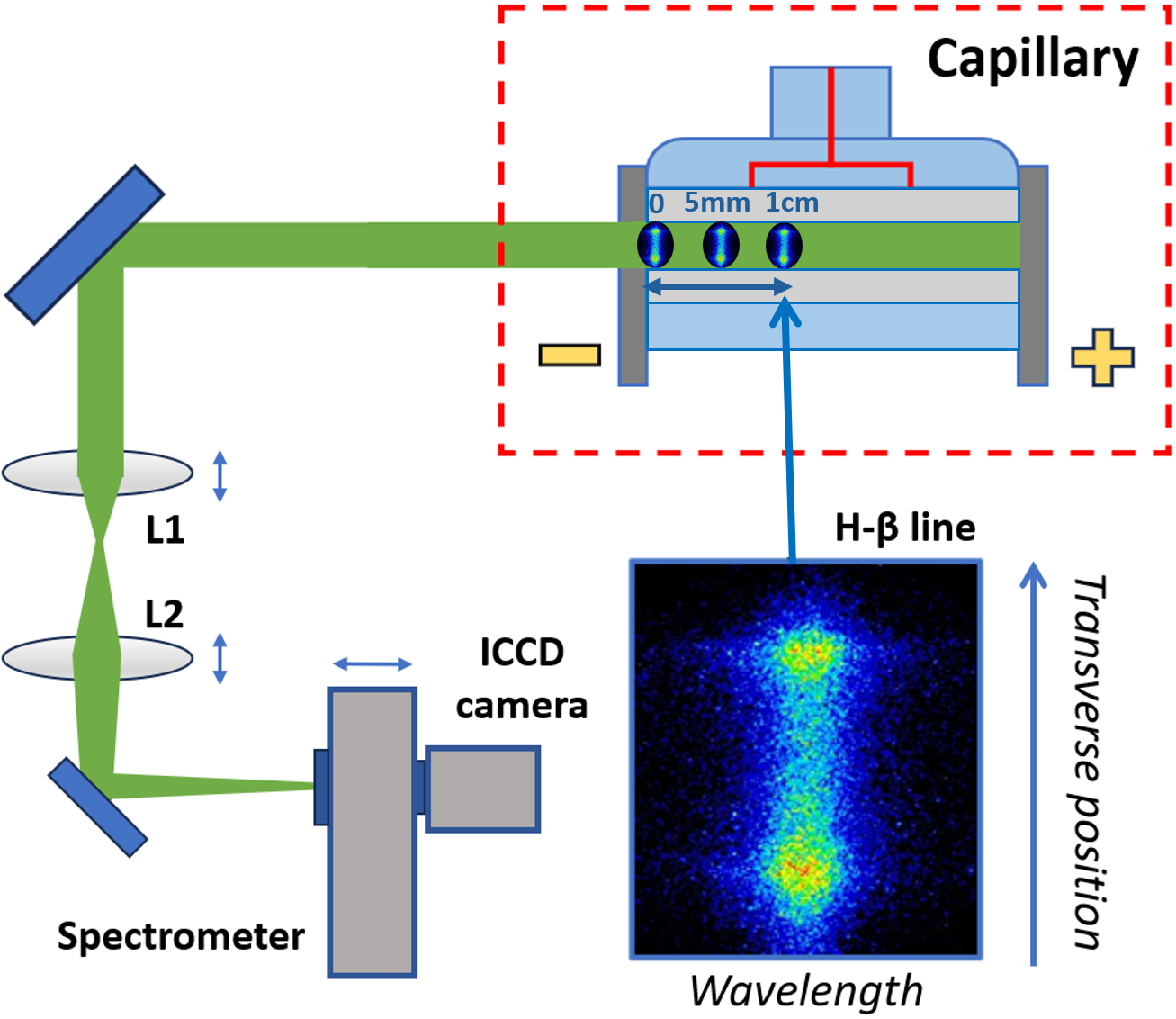}
\caption{Schematic representation of the transverse Stark broadening diagnostics. By shifting the telescopic system longitudinally, different slices of the plasma channel are acquired, performing a longitudinal scan of the transverse plasma density distribution.}
\label{transverse_stark}
\end{figure*}

As shown in Fig.\ref{transverse_stark}, acquired spectral images allow to reconstruct the transverse (vertical) plasma density distribution corresponding to a specific slice of the plasma channel, which is imaged onto the ICCD camera detector by the beam transport line.
%Spectral images acquired by the ICCD camera allow to reconstruct the transverse plasma density profile inside the capillary channel and.
The camera acquisition time is synchronized with the HV pulser to characterize the plasma density temporal evolution during the electrical discharge.
Furthermore, by moving the telescopic system and the spectrometer along the beam transport line, it is possible to scan the plasma channel in the longitudinal direction, analyzing the transverse plasma distribution from different channel slices.

Since the Stark broadening method specifically relies on the broadening of hydrogen spectral lines, plasma density measurements are performed with pure H$_2$ gas, delivered by a hydrogen generator which operates in pulsed injection regime.
%In order to optimize the reliability of plasma density measurements, pure hydrogen gas is employed by means of a hydrogen generator.
%However, due to technical limitations of the hydrogen generator, which is not feasible for continuous flow injection, plasma density measurements are performed with pulsed gas injection at 1 Hz, controlled by means of an electro-mechanical valve.
Therefore, while high repetition rate tests are carried out with the mixture of N$_2$ (95\%) and H$_2$ (5\%) gas, injected in continuous flow, spectroscopic measurements are performed with pure H$_2$, injected at 1 Hz by means of an electro-mechanical valve.

Additionally, spectral images are acquired with a 150 grooves/mm grating to analyze the entire plasma emission spectrum in the visible range, which gives information on the ion and atomic species present in the plasma.
Furthermore, by measuring the relative intensity of Balmer H$_{\alpha}$ and H$_{\beta}$ spectral lines, it is possible to obtain an estimate on the electron plasma temperature, according to \cite{Griem2}:

\begin{equation}\label{Temperature_meas}
    R = \frac{I_{H_\alpha}}{I_{H_\beta}} = \frac{\omega_{H_\alpha} A_{H_\alpha} g_{H_\alpha}}{\omega_{H_\beta} A_{H_\beta} g_{H_\beta}} exp\left(-\frac{E_{H_\alpha}-E_{H_\beta}}{k_B T_e}\right)
\end{equation}

in which $I$, $\omega$, $g$ and $A$ are the intensity, frequency, degeneracy and strength of H$_{\alpha}$ and H$_{\beta}$ optical transitions, while $E$ represents the corresponding excitation energy, all of which are retrieved by the NIST Atomic Spectra Database \cite{ASD}.
Due to the relatively small differences between excitation energies and theoretical uncertainties, the ratio $R$ of the spectral line intensities related to the same hydrogen atom does not lead to precise temperature measurements \cite{Griem2}, therefore this method is only adopted for an experimental comparison with the theoretical estimate of the plasma temperature from Eq.\ref{Bobrova}.

% \begin{figure}[!ht]
% \centering
% \includegraphics[width=0.9\linewidth]{Images/Transverse Stark scheme.png}
% \caption{Scheme of the Stark broadening method to characterize the transverse plasma distribution}
% \label{Transverse Stark scheme}
% \end{figure}

%\FloatBarrier

\section*{Results and discussion}

\subsection*{Electrical diagnosis}
\begin{figure}[h!]
    \centering
    \begin{subfigure}[h!]{0.47\textwidth}
        \centering
        \includegraphics[width=\textwidth]{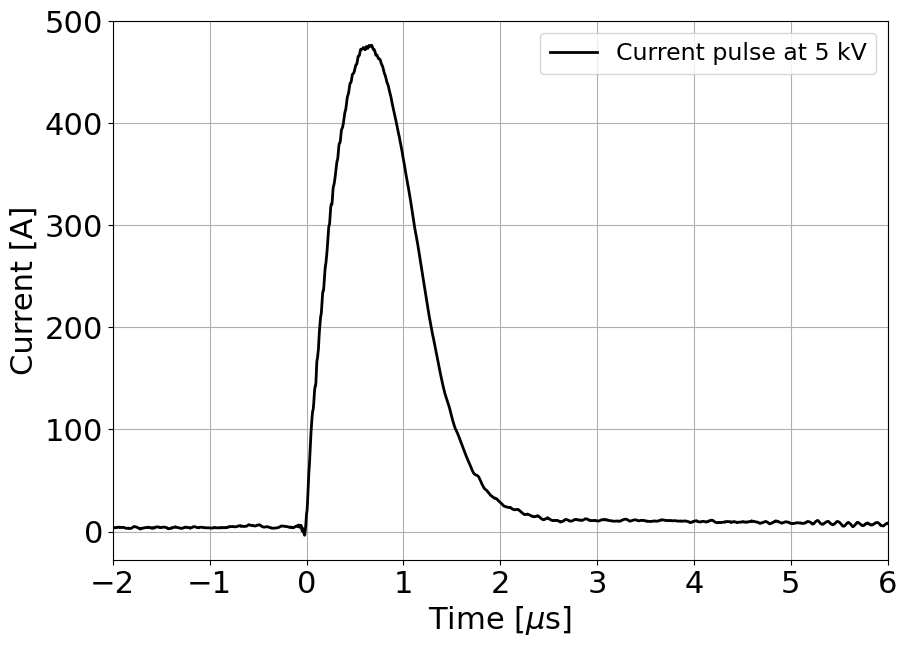}
        \caption{Discharge current waveform acquired by the oscilloscope}
        \label{wvfrm}
    \end{subfigure}
    \hfill
    \begin{subfigure}[h!]{0.51\textwidth}
        \centering
        \includegraphics[width=\linewidth]{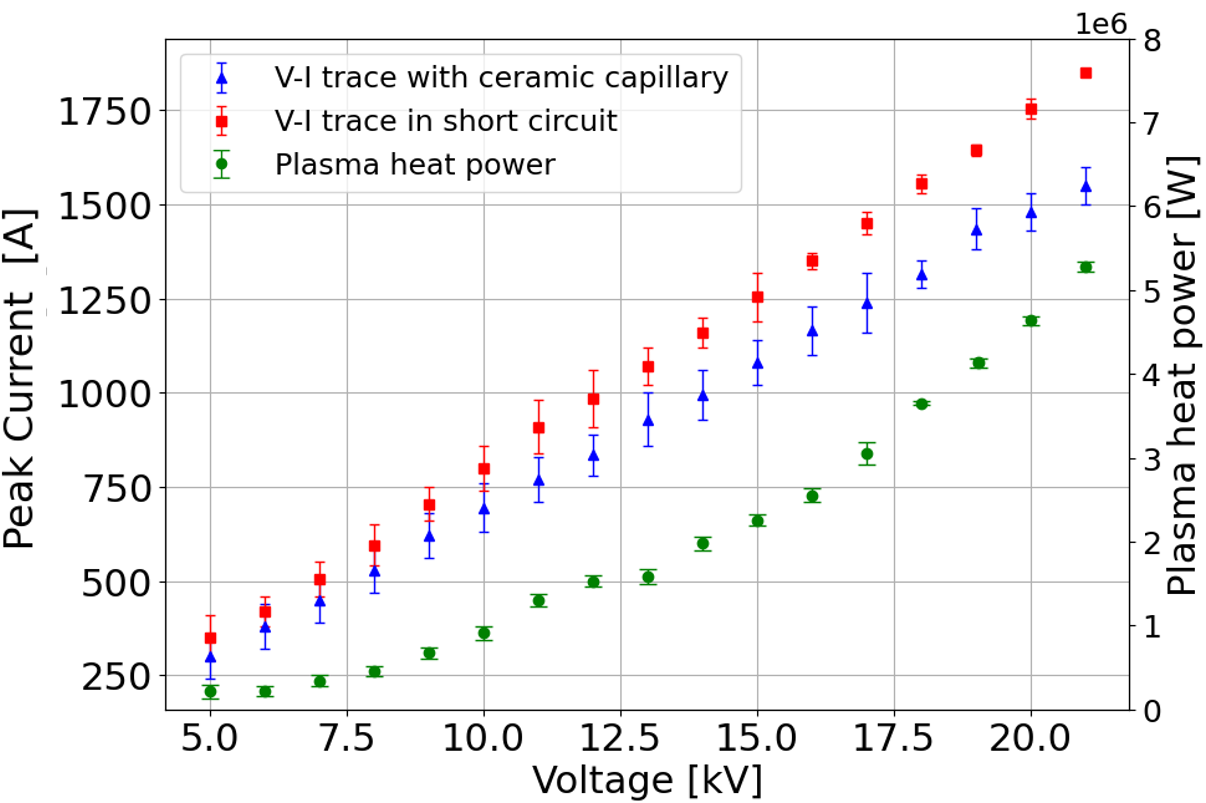}
        \caption{Voltage-Current traces and instantaneous plasma heat power}
        \label{V_I}
    \end{subfigure}
    \caption{(a) Current pulse waveform produced with 5 kV pulses applied to the 3 cm-long 2 mm-diameter ceramic capillary. The FWHM pulse duration is around 1 $\mu$s. (b) Voltage-Current traces of the HV electrical pulser, measured in short circuit (red squares) and with the ceramic capillary (blue triangles). The lower peak current reached with the capillary is due to the plasma resistance. Each point is computed as the standard deviation over 50 measurements. In addition, the instantaneous heat power produced by the plasma discharge (green circles) is reported.}
    \label{I and V_I}
\end{figure}

An electrical diagnosis is performed on the HV circuit, providing information on the discharge current and the plasma resistance and heat power.
%The typical waveform of a current pulse, produced by a HV capillary discharge and acquired by the oscilloscope, is shown in Fig.\ref{wvfrm}.
%The reported discharge current waveform is obtained by applying 5 kV pulses to the 3 cm-long 2 mm-diameter ceramic capillary, ionizing around 50 mbar of N$_2$-H$_2$ mixture.
The waveform of the current pulse produced by the circuit is acquired by the oscilloscope, as shown in Fig.\ref{wvfrm}, while varying the voltage applied by the HV generator, thus allowing to retrieve the characteristic voltage-current curve of the pulser circuit.
The reported discharge current waveform is obtained by applying 5 kV pulses to the 3 cm-long 2 mm-diameter ceramic capillary, ionizing around 20 mbar of N$_2$-H$_2$ mixture.
As shown in Fig.\ref{V_I}, V-I curves are determined for a short circuit configuration and with the ceramic capillary, with maximum peak current of 1850 A and 1550 A respectively.
The lower current produced with the capillary is due to the electrical resistance of the plasma channel, which increases with the capillary length.

By comparing the V-I traces in the two configurations, it is possible to determine the plasma resistance $R_{p}$ as the difference between the total resistance $R_{tot}$ (circuit and capillary) and the circuit resistance in short circuit configuration $R_{sc}$ (without the capillary):

\begin{equation}
    R_{p} = R_{tot} - R_{sc} = V/I_{plasma}-V/I_{sc}
\end{equation}

given $R_{tot}$ and $R_{sc}$ by the ratio between the applied voltage and the peak current measured with the capillary and in short circuit respectively.
In particular, in the experimental range of 5-21 kV, the plasma resistance spans from 1.5 to 2.2 $\Omega$.
Moreover, by multiplying the plasma resistance with the measured current intensity, it is possible to determine the instantaneous heat power produced by the plasma discharge through Ohmic heating, according to \cite{PhysRevE.100.053202}:

\begin{equation}\label{Joule}
P = R_p I_p^{2}
\end{equation}

The instantaneous heat power, reported in Fig.\ref{V_I} as a function of the applied voltage, spans from 100 kW to 5 MW in the experimental range of 5-21 kV.
Therefore, considering the discharge pulse duration of 1 $\mu$s FWHM, the heat produced by a single plasma discharge ranges from 100 mJ to 5 J.
As a result, the average heat power deposited onto the capillary walls during continuous operation at low voltage (e.g. 5 kV) in the range of 10-100 Hz turns out to be around 1-10 W.

%\newpage
\subsection*{Preliminary characterization}
\begin{figure}[h!]
\centering
\includegraphics[width=0.85\linewidth]{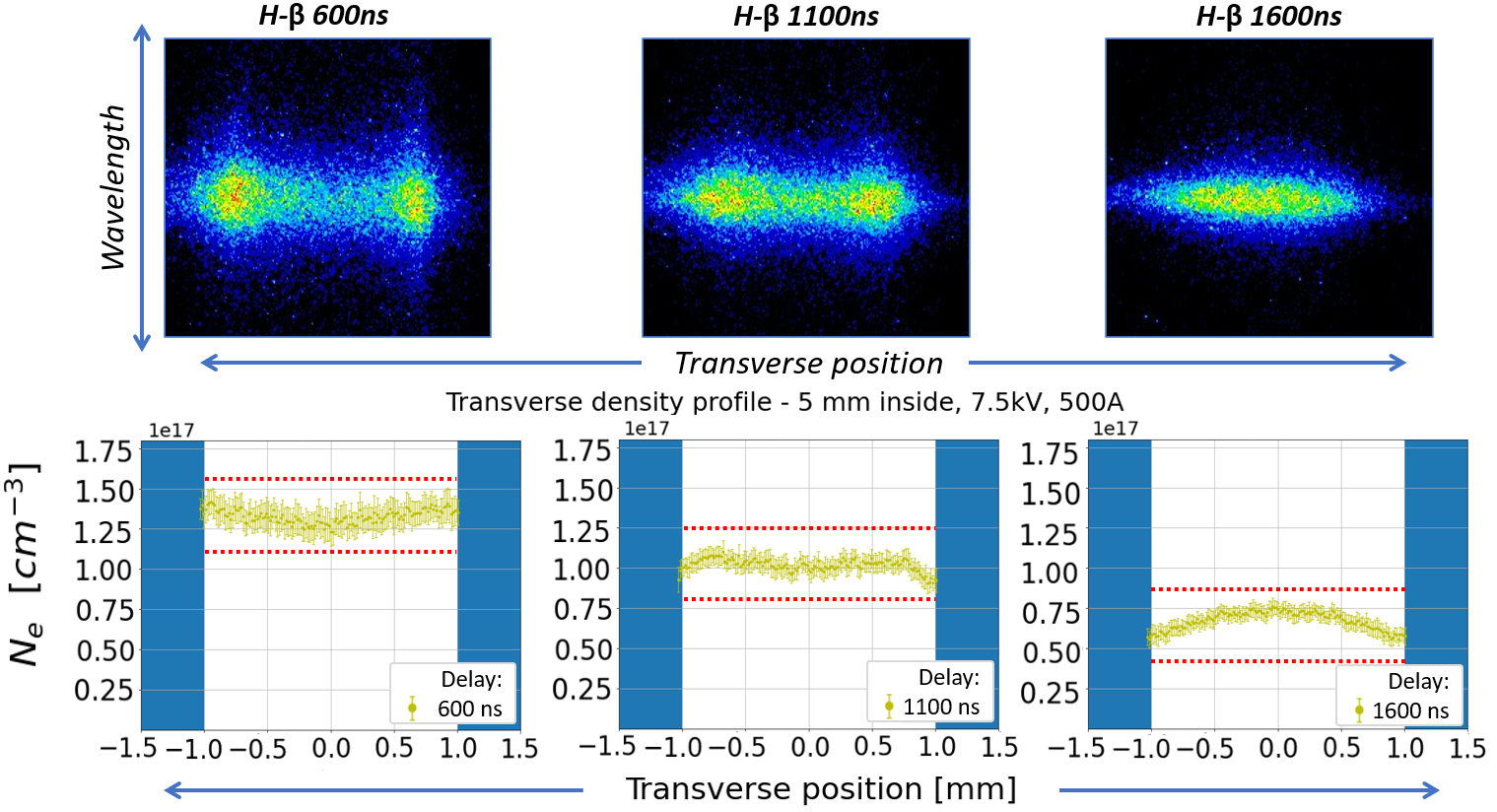}
\caption{(Top) H\textsubscript{$\beta$} spectral images and (Bottom) transverse plasma density profiles, measured from the plasma channel slice located 5 mm inside the capillary at different delays with respect to the onset of the HV discharge.
%(Right) Time evolution of the plasma density averaged over the transverse profiles, measured in three different plasma channel slices: 1 cm inside (red), 5 mm inside (yellow) and at the exit (blue).
}
\label{Slices comparison}
\end{figure}

Before high repetition rate tests, a first characterization of the Shapal capillary is performed. 7.5 kV voltage pulses are delivered to the capillary electrodes, ionizing around 20 mbar of pure H$_2$ and producing plasma discharges at 1 Hz, with 500 A peak current.
The telescopic system of the beam transport line is shifted longitudinally over 1 cm to scan the plasma channel and acquire the plasma-emitted light from different transverse slices, in particular from the capillary exit to 5 mm and 1 cm inside the capillary channel, as schematized in Fig.\ref{transverse_stark}.
Hydrogen Balmer lines are then analyzed to retrieve the transverse plasma density distribution in the different slices. %and towards the exit.
Results from Stark broadening method, reported in Fig.~\ref{Slices comparison}, show the evolution of the transverse plasma density profile 5 mm from the capillary exit, at different delays with respect to the onset of the HV discharge.

During the plasma formation, the density distribution is characterized by a hollow profile with a depth of 18\% with respect to the near-wall density (around $\pm$ 1 mm), as observed at a delay of 600 ns at the density peak of around 1.3$\times$$10^{17}$\ cm$^{-3}$\ . In the recombination phase, after the discharge is over (delay beyond 1000 ns), the transverse distribution turns into a gaussian-shaped profile.
Measured error bars represent the standard deviation calculated by acquiring 50 spectral images.

\begin{figure}[h!]
\centering
\includegraphics[width=0.56\linewidth]{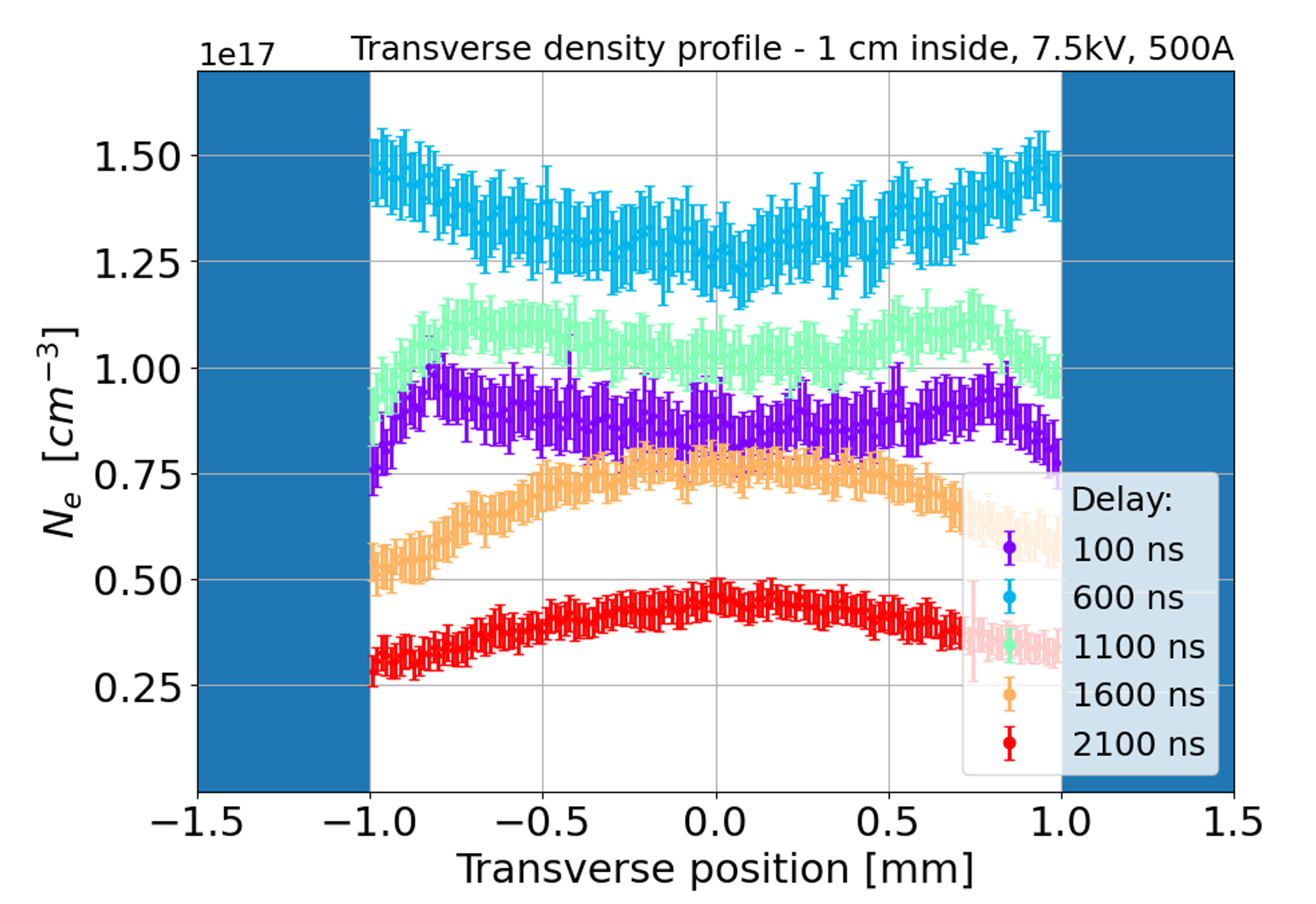}
\caption{Transverse plasma density profiles, measured 1 cm inside the capillary at different delays after the discharge onset.}
\label{Transv prof s0}
\end{figure}

Observed transverse density profiles are in good agreement with theoretical models\cite{Bobrova}, showing that, given a uniform plasma pressure, the radial temperature gradient, established between the channel axis and the capillary walls due to heat transfer with the capillary, results in a hollow density profile with an on-axis minimum and a density maximum towards the channel walls.
The same behaviour is observed with the transverse slice located 1 cm inside the plasma channel, as reported in Fig.~\ref{Transv prof s0}.

% for the transverse profiles measured 5 mm inside the channel.
%and the time evolution of the average plasma density for the three slices.
%Plasma density profiles from the three channel slices do not differ significantly, because the depth of field of the optical line is not short enough to acquire the plasma light coming from a thin slice of the channel, but actually a large portion of the channel is collected.

\begin{figure}[h]
\centering
\includegraphics[width=0.69\linewidth]{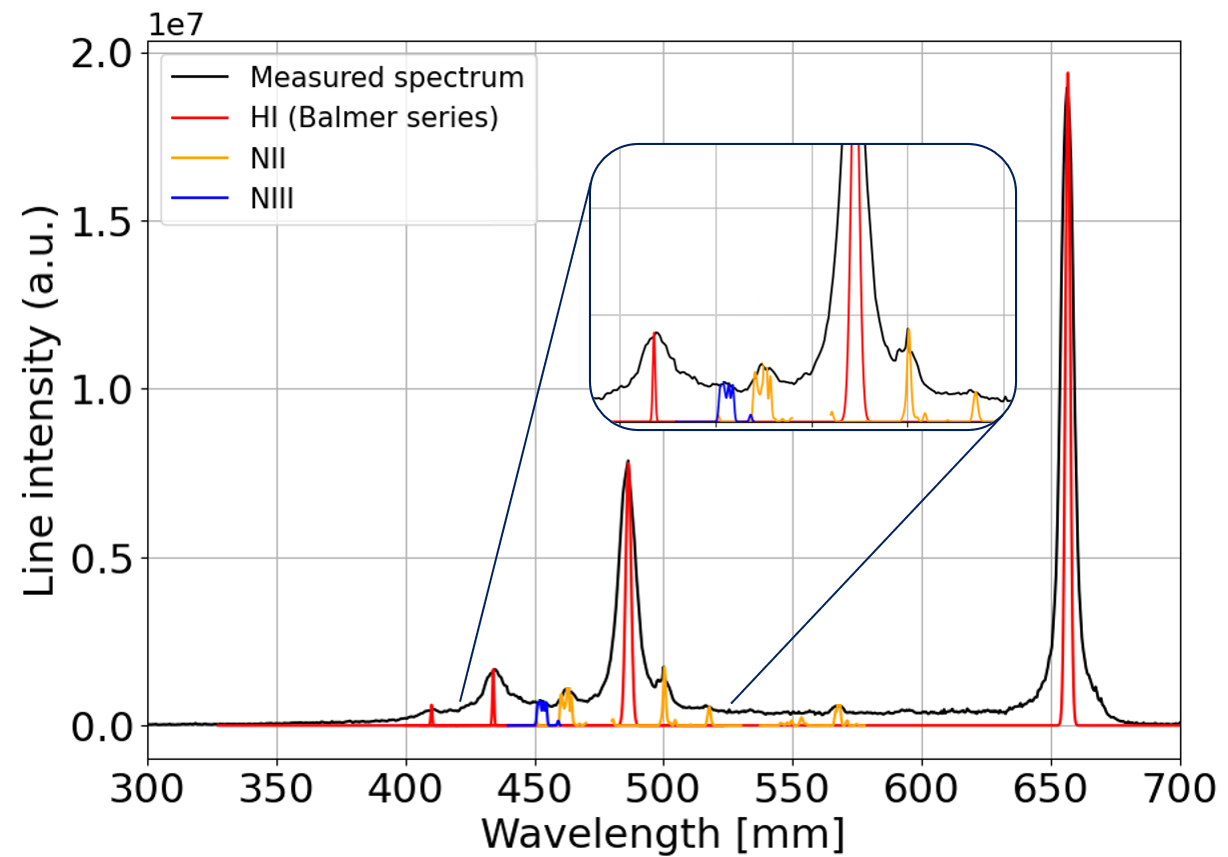}
\caption{Plasma emission spectrum, acquired 100 ns after the discharge onset from the plasma channel slice located 1cm inside the capillary and integrated along the transverse profile. The measured spectrum (black line) is benchmarked with the NIST LIBS Database spectrum tabulated for a hydrogen plasma with few percent nitrogen and density and temperature of 10$^{17}$ cm$^{-3}$ and 3 eV respectively. In particular, the Database spectrum shows the hydrogen lines of the Balmer series (red line) and also emission lines of $N^+$ (NII, orange line) and $N^{2+}$ (NIII, blue line).}
\label{Spectrum}
\end{figure}

\newpage
Regarding the plasma composition, Fig.\ref{Spectrum} reports the plasma emission spectrum acquired by the 150 grooves/mm spectrometer grating and integrated over the transverse slice located 1 cm inside the capillary.
The spectrum is acquired at a delay of 100 ns and shows the hydrogen lines of the Balmer series, together with emission lines of $N^+$ (centered at 464, 500, 517 and 568 nm) and $N^{2+}$ (452 nm), thus showing the trace presence of nitrogen in the plasma channel, due to residual air in the gas injection pipe, which then vanishes at higher delays.
Concerning the plasma temperature, Eq.\ref{Bobrova} yields a temperature of around 3 eV during the plasma formation and a peak of 4.3 eV in correspondence of the 500 A peak current.
Such result is confirmed by the relative intensities of H\textsubscript{$\alpha$} and H\textsubscript{$\beta$} spectral lines, measured from Fig.\ref{Spectrum} and inserted into Eq.\ref{Temperature_meas}.

The acquired emission spectra and the measured plasma electron density and temperature are benchmarked with NIST LIBS Database spectra \cite{LIBS}, tabulated for a hydrogen plasma doped with few percent of nitrogen and with electron density and temperature of 10$^{17}$ cm$^{-3}$ and 3-4 eV respectively.
As shown in Fig.\ref{Spectrum}, the acquired spectrum is well overlapped with LIBS tabulated lines with the plasma parameters obtained through spectroscopic analysis, except for the broadening of Balmer lines, which is not taken into account into NIST spectrum.

\subsection*{High repetition rate operation}

\begin{figure}[h!]
    \centering
    \begin{subfigure}[h!]{0.5\textwidth}
        \centering
        \includegraphics[width=\textwidth]{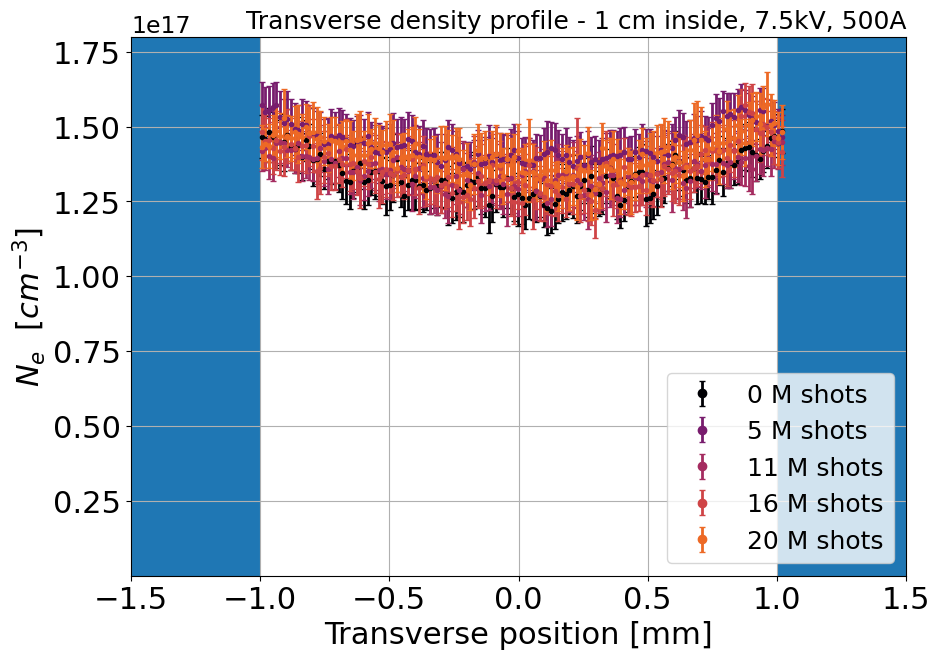}
        \caption{}
        \label{Density evolution 1}
    \end{subfigure}
    \begin{subfigure}[h!]{0.48\textwidth}
        \centering
        \includegraphics[width=\textwidth]{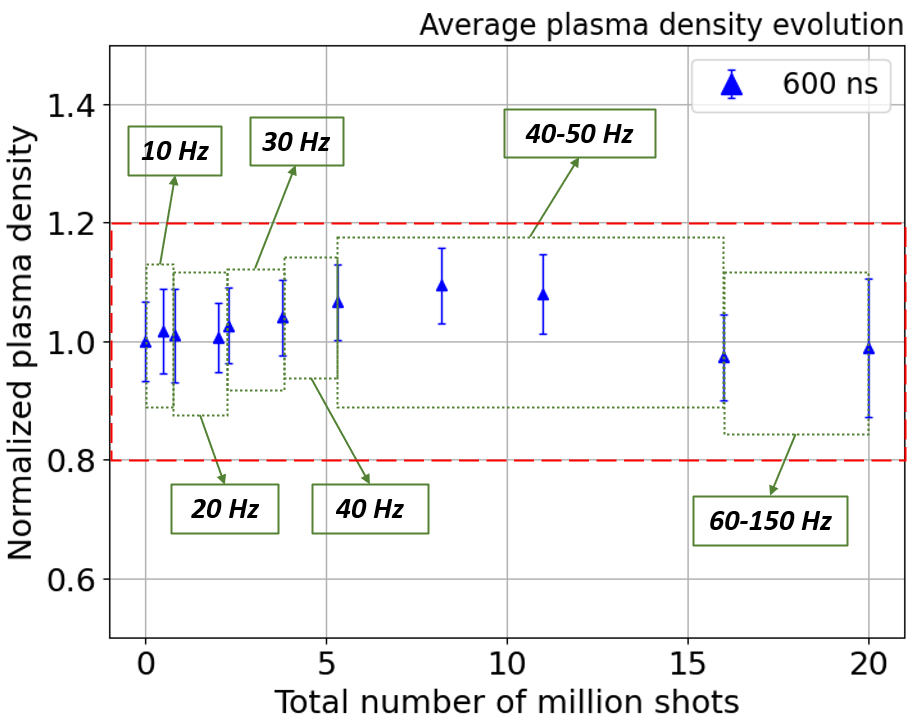}
        \caption{}
        \label{Density evolution 2}
    \end{subfigure}
    \caption{(a) Transverse density profiles and (b) average plasma density, measured 1 cm inside the plasma channel at a delay of 600 ns, after different number of shots at 10-150 Hz up to 20 million shots.}
    \label{Density evolution}
\end{figure}

%After the initial characterization at 1 Hz
High repetition rate tests are performed in the range of 10-150 Hz, using the N$_2$-H$_2$ mixture in continuous flow regime, injected at 80-100 mbar.
%, corresponding to around 10 mbar inside the capillary.
5 kV voltage pulses are delivered to the electrodes, producing 400 A peak current plasma discharges, up to a total amount of 20 million shots.
%with 40 ns time jitter, measured by an oscilloscope without using laser pulses for the discharge stabilization.
During the experimental campaign, laser spot imaging and plasma density measurements are performed regularly to monitor any modification both in the capillary walls and the plasma density distribution.
In particular, after a given number of shots, plasma density measurements are performed using 7.5 kV pulses to ionize pure hydrogen in pulsed flow regime, as for the preliminary characterization, and taking as a reference the transverse plasma density profiles acquired 1 cm inside the channel at the density peak (600 ns after the HV discharge onset). % and acquired 1500 ns after the 500 A peak current discharge.
Higher voltage pulses are applied, compared to 5 kV pulses in high repetition rate tests, in order to maximize the plasma stability and the output signal, thus minimizing the error in density measurements.
Results obtained with Stark broadening method are reported in Fig.~\ref{Density evolution 1}.
Transverse density profiles, measured in the same experimental conditions after a different number of discharges, are well overlapped within the error bars, which are computed by acquiring 50 images for each measure.
Furthermore, as depicted in Fig.~\ref{Density evolution 2}, the plasma density, averaged over the transverse profile and normalized to the value obtained in the preliminary characterization, remains approximately constant during the entire experimental campaign at 10-150 Hz. % at around 1.3$\times$$10^{17}$\ cm$^{-3}$, 

% \begin{figure}[h!]
% \centering
% \includegraphics[width=0.51\linewidth]{Images/Progressive profiles inferno.png}
% \includegraphics[width=0.47\linewidth]{Images/t evolution 20M normalized 2.png}
% \caption{Transverse density profiles (left) and average plasma density (right), measured 1 cm inside the plasma channel at a delay of 600 ns, after different number of shots at 10-150 Hz.% (Right) Evolution of the plasma density, averaged over the transverse profile, after . %Reported density values are normalized to the first measurement performed before high repetition rate tests.
% }
% \label{Density evolution}
% \end{figure}

\begin{figure}[h!]
\centering
\includegraphics[width=0.72\linewidth]{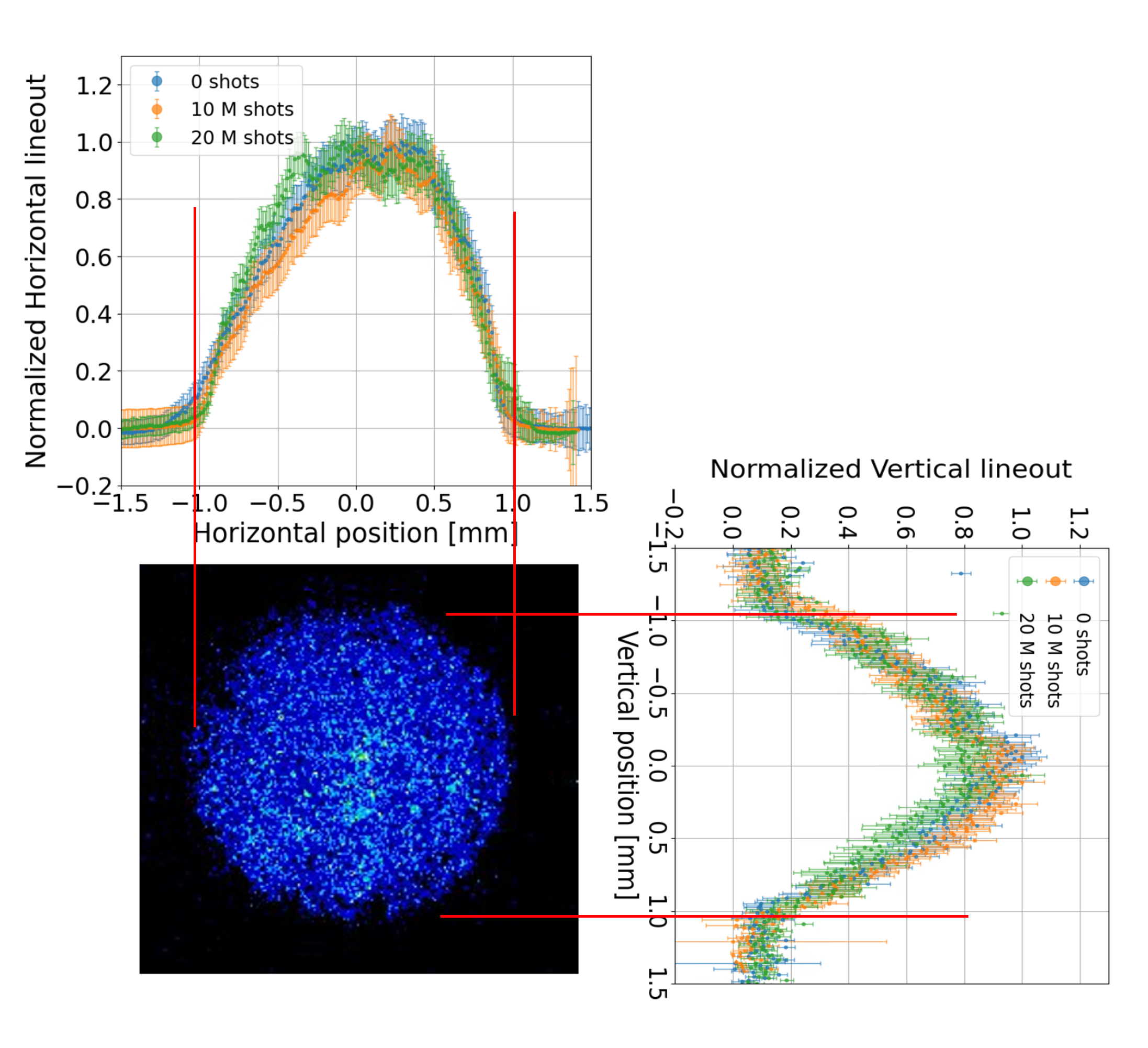}
\caption{Laser spot vertical and horizontal lineouts progressively measured up to 20 million shots at 10-150 Hz}
\label{Laser spots}
\end{figure}

\newpage
Regarding the capillary channel cross section, the analysis of laser spot images is reported in Fig.~\ref{Laser spots}. Horizontal and vertical laser spot lineouts, measured during the experimental campaign, are well overlapped within the error bars, which are computed by acquiring 50 images for each measurement.
Therefore, no significant alteration is observed in the overall channel profile.

\begin{figure}[h!]
\centering
\includegraphics[width=0.55\linewidth]{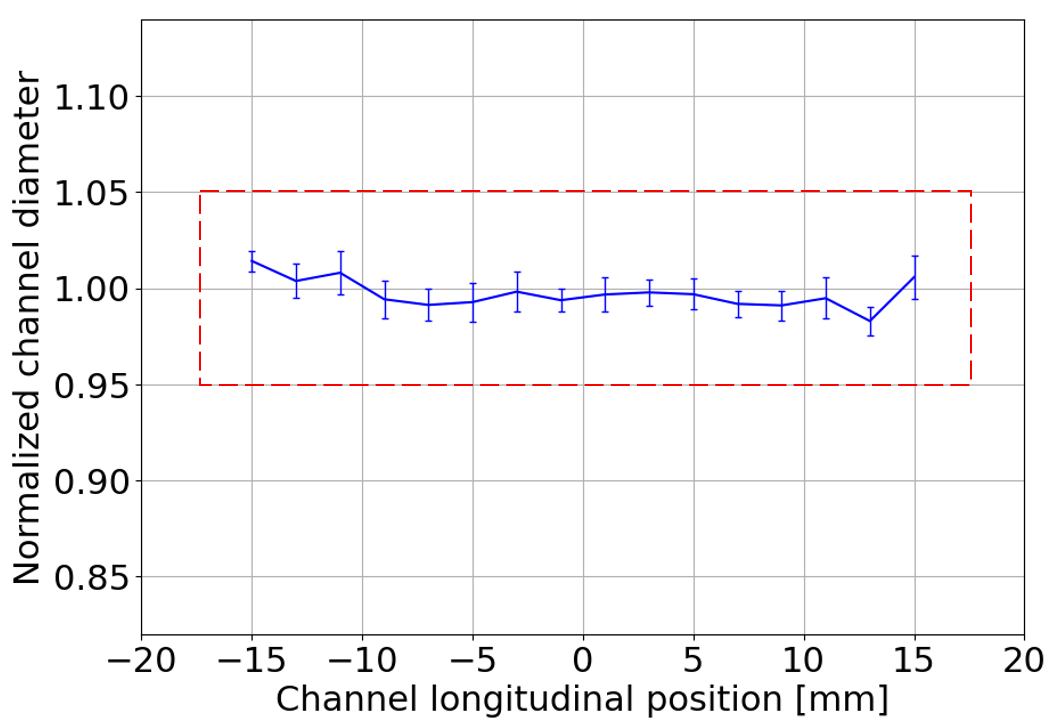}
\caption{Capillary channel diameter, measured by the stereomicroscope along the capillary axis after 20 million shots at 10-150 Hz, normalized with respect to the first measurement performed before the experimental campaign.
%(Right) Microscopic images of the capillary extremities before (left) and after (right) high repetition rate test. The right side connected to the anode (bottom) is particularly affected by steel deposition, caused by electrode sputtering.
}
\label{Average diameter}
\end{figure}

In addition, concerning the microscopic analysis, Fig.~\ref{Average diameter} reports the channel diameter, measured along the capillary after 20 million discharges and normalized to the channel profile acquired before the experimental campaign.
For each longitudinal position, the reported diameter is obtained by measuring the channel diameter both in the horizontal and vertical directions and computing the average, in order to take into account possible deformations that would determine a transverse elliptical shape. Measured error bars result from 200 acquired images.
As a result, microscopic analysis confirms that the capillary channel is well preserved at the end of the experimental campaign.

In conclusion, plasma measurements highlight the stability of the density distribution inside the capillary during the experimental campaign, while the laser spot imaging and the microscopic analysis show that the capillary integrity is preserved as well. Hence experimental results demonstrate exhaustively the suitability of the Shapal-Macor capillary for long-term plasma discharges operation at high repetition rate (up to 150 Hz).

% \begin{figure}[t]
% \centering
% \includegraphics[width=0.49\linewidth]{Images/Microscope images.png}
% \caption{Microscopic images of the capillary extremities before (left) and after (right) 15 million shots at 10-50 Hz. The right side connected to the anode (bottom) is particularly affected by steel deposition, caused by sputtering}
% \label{Erosion}
% \end{figure}

%\newpage
\subsection*{Heat transfer numerical simulations}
%\subsection{Numerical simulations}
A numerical analysis is carried out to evaluate the heat transfer inside the entire plasma source during high repetition rate operation. % an operative limit of the Shapal capillary, given by the high repetition rate overheating.
%given by the overheating caused by the thermal load delivered by HV discharges onto the capillary walls.
First, the instantaneous heat power produced by a single plasma discharge through Ohmic heating is considered as:

\begin{equation}\label{Joule}
P(t) = R_p(t) I_p(t)^{2}
\end{equation}

%in which I$_p$ and R$_p$ are the discharge current  and the plasma resistance respectively.
The discharge current $I_p$ is directly measured by the oscilloscope during the experimental campaign, while the resistance of the plasma channel  $R_p$ is estimated through Ohm's law\cite{Spitzer1956PhysicsOF}:

\begin{equation}\label{Ohm}
R_p(t) = \rho_{tot}(t) \frac{L}{\pi r^2},
\end{equation}

in which the length $L$ and radius $r$ of the capillary channel are respectively 3 cm and 1 mm, and the plasma resistivity $\rho_{tot}$ is computed as\cite{ANANIA2014193, ANANIA2016254}:

\begin{equation}\label{Spitzer}
\rho_{tot}(t) = \frac{m_{e}}{n_{e}(t) e^2} (\nu_{ei}(t) + \nu_{ae}(t))
\end{equation}

Electron-ion and atom-electron collision frequencies are in turn determined as \cite{Huba2004NRLPF}:

\begin{equation}\label{nuae}
\nu_{ei}(t) = \frac{4}{3} \sqrt{\frac{2\pi}{m_{e}}}\frac{e^4 n_{e}(t) \ln{\lambda_{ei}(t)}}{(4\pi \epsilon_{0})^2 (k_{B} T_{e}(t))^{3/2}}, \quad \nu_{ae}(t) = \frac{\pi r^2_{a} P_{0}}{\sqrt{m_{e} k_{B} T_{e}(t)}}
\end{equation}
%\end{equation}

%\begin{equation}\label{nuei}

in which $r_{a}$ and $P_0$ are the atomic radius and the gas pressure, while the Coulomb logarithm $\ln{\lambda_{ei}}$ is given by \cite{Huba2004NRLPF}:

\begin{equation}\label{logCoulomb}
\ln{\lambda_{ei}(t)} = \ln{\left[\frac{3}{2\sqrt{2\pi}}\frac{ (4\pi \epsilon_{0})^{3/2} (k_{B} T_{e}(t))^{3/2}}{e^3 [n_e(t)]^{1/2}} \right]}
\end{equation}

\begin{figure}[h!]
    \centering
    \begin{subfigure}[h!]{0.49\textwidth}
        \centering
        \includegraphics[width=\textwidth]{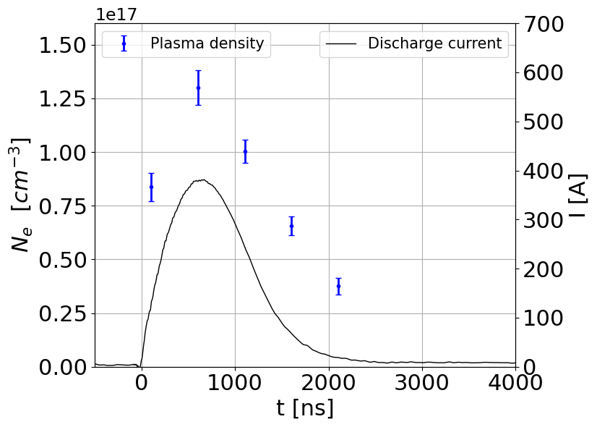}
        \caption{}
        \label{Ne input}
    \end{subfigure}
    \begin{subfigure}[h!]{0.49\textwidth}
        \centering
        \includegraphics[width=0.95\textwidth]{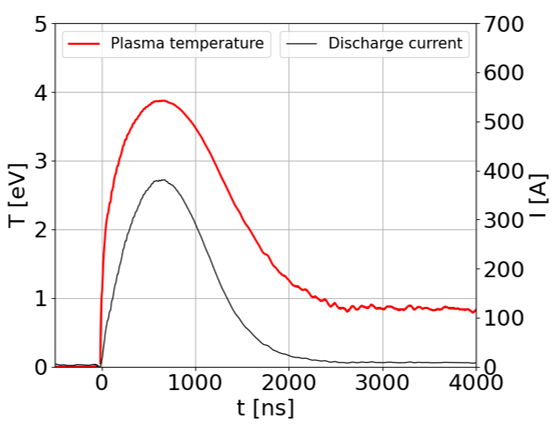}
        \caption{}
        \label{Te input}
    \end{subfigure}
    \caption{Temporal profiles of the plasma density (a) and temperature (b), respectively determined by transverse Stark broadening method and Eq.\ref{Bobrova}.}
    \label{Analytical input}
\end{figure}

%The temporal profiles of the plasma density, reported in Fig.~\ref{Transv prof s0}, and the plasma temperature, obtained through [\ref{Bobrova}],
%Temporal profiles of the plasma density, obtained by averaging the measured transverse profiles in Fig.~\ref{Transv prof s0}, and the plasma temperature, recovered by Eq. (\ref{Bobrova}), are reported in Fig.~\ref{Analytical input} and inserted into Eq (\ref{nuae}, \ref{logCoulomb}) to compute the Coulomb logarithm and collision frequencies and, in turn, recover the plasma resistivity and the resistance of the plasma channel.
In order to compute the Coulomb logarithm and the collision frequencies and, in turn, retrieve the plasma resistivity and the plasma channel resistance, the temporal profiles of the plasma density and the plasma temperature, depicted in Fig.~\ref{Analytical input}, are inserted into Eq (\ref{Spitzer}, \ref{nuae}, \ref{logCoulomb}). In particular, the density temporal profile is obtained by averaging the measured transverse profiles from Fig.~\ref{Transv prof s0}, while the temperature temporal profile is computed by implementing the measured discharge current into Eq.\ref{Bobrova}.

\begin{figure}[h!]
    \centering
    \begin{subfigure}[h!]{0.33\textwidth}
        \centering
        \includegraphics[width=0.99\textwidth]{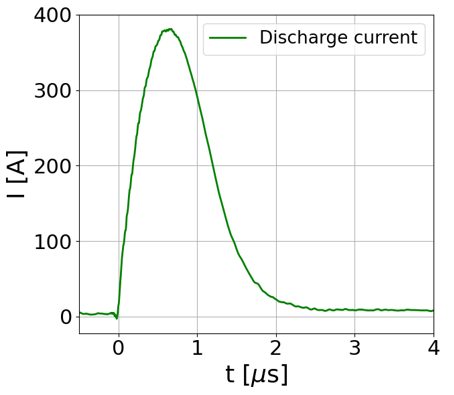}
        \caption{}
        \label{I output}
    \end{subfigure}
    \begin{subfigure}[h!]{0.33\textwidth}
        \centering
        \includegraphics[width=0.95\textwidth]{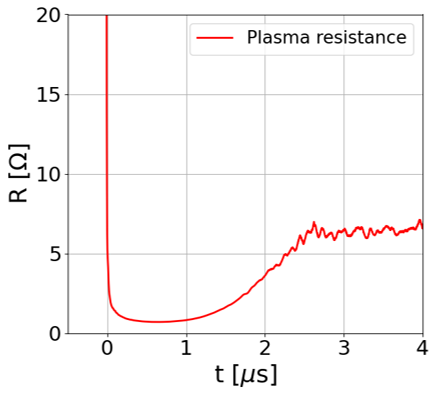}
        \caption{}
        \label{R output}
    \end{subfigure}
    \begin{subfigure}[h!]{0.33\textwidth}
        \centering
        \includegraphics[width=0.99\textwidth]{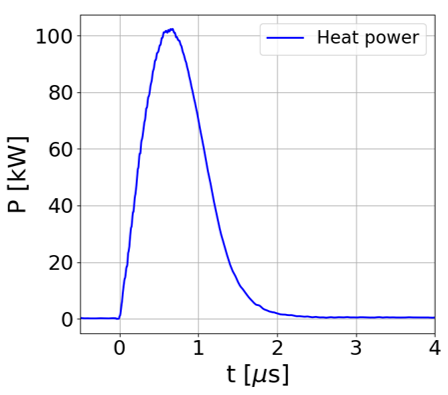}
        \caption{}
        \label{P output}
    \end{subfigure}
    \caption{(a) Discharge current waveform, acquired during high repetition rate tests. (b) Resistance of the plasma channel. (c) Resulting instantaneous heat power produced by a single plasma discharge.}
    \label{Analytical results}
\end{figure}

%density measured in the experimental tests, with peak values of around 4 eV and $10^{17}$\ cm$^{-3}$\ respectively. %, thus resulting in a Spitzer resistivity of around 5.2 $\Omega\cdot$m and a plasma resistance of 0.5 $\Omega$.
Analytical results for the plasma resistance and the instantaneous heat power are reported in Fig.~\ref{Analytical results}, together with the discharge current waveform measured during high repetition rate tests.
In particular, the obtained plasma resistance and instantaneous heat power are in agreement with experimental results achieved through the electrical diagnosis of the HV circuit and the capillary discharge, as shown in Fig.\ref{V_I}.

% \begin{figure}[h!]
% \centering
% \includegraphics[width=0.99\linewidth]{Images/Analytical output I-R-P 3.png}
% \caption{}
% \label{Analytical results}
% \end{figure}

Finally, by integrating the instantaneous heat power over the plasma discharge duration, an energy per pulse of $\approx$ 100 mJ is obtained.
%As a result, the energy produced through Ohmic heating by a single discharge is around 60 mJ.
Considering repetition rate operation from tens of Hz up to the kHz range, the average heat power deposited onto the capillary walls spans from few Watts (10-50 Hz) to 100 W. % (up to 100 kHz).
Given these reference values, 3D numerical simulations are performed with COMSOL Multiphysics \cite{COMSOL} to analyze the capillary overheating and the heat removal inside the plasma source. Relying on Fourier's law of heat conduction, the temperature gradient inside the whole source is computed, given the thermal conductivity \textit{k} of the different components and the heat flux \textit{q}, estimated through the analytical model previously described:

\begin{equation}\label{Fourier}
\Vec{q} = -k(T)\nabla T
\end{equation}

\begin{figure}[h!]
    \centering
    \begin{subfigure}[h!]{0.49\textwidth}
        \centering
        \includegraphics[width=0.9\textwidth]{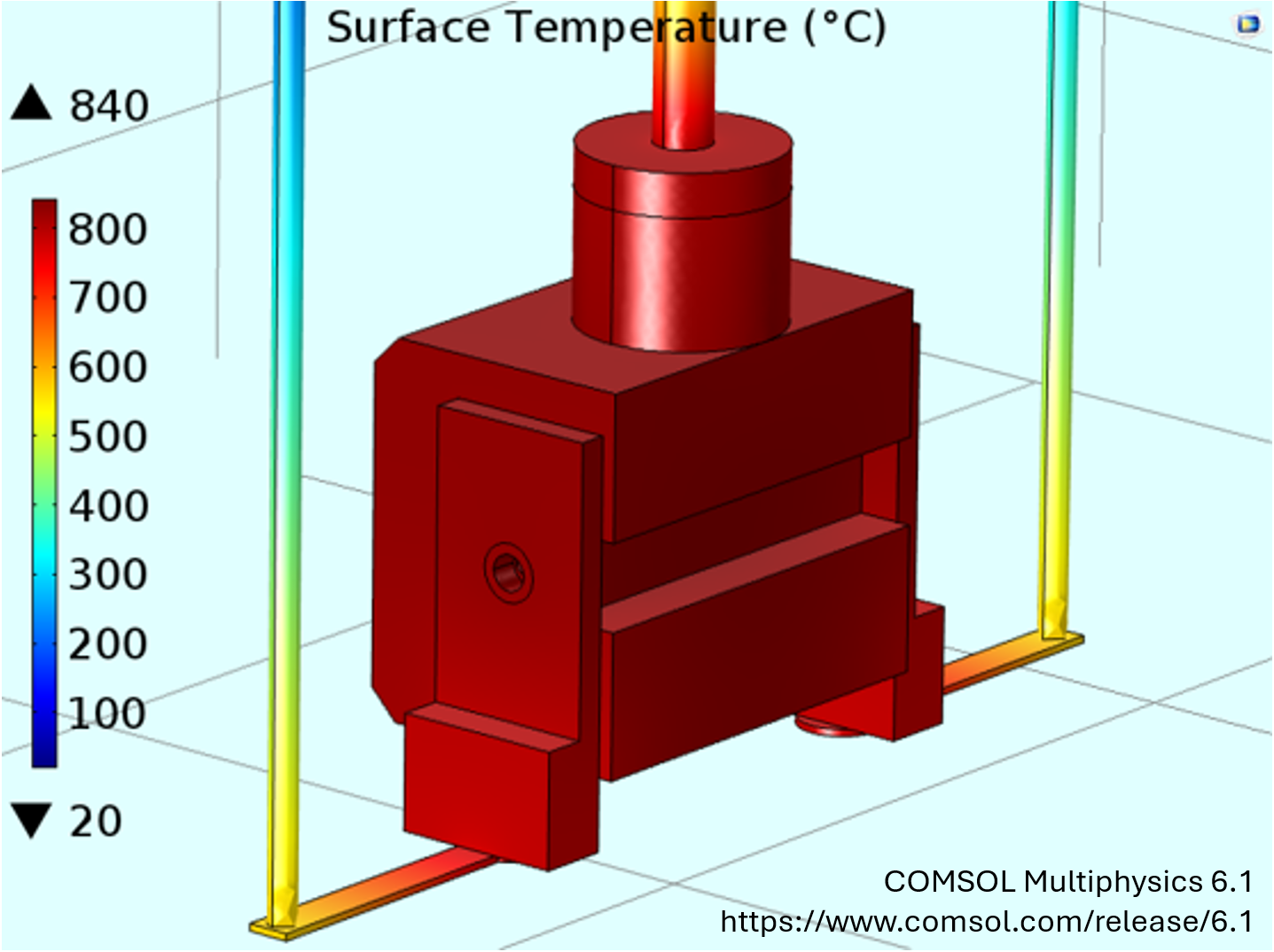}
        \caption{}
        \label{3D view}
    \end{subfigure}
    \hfill
    \begin{subfigure}[h!]{0.49\textwidth}
        \centering
        \includegraphics[width=\textwidth]{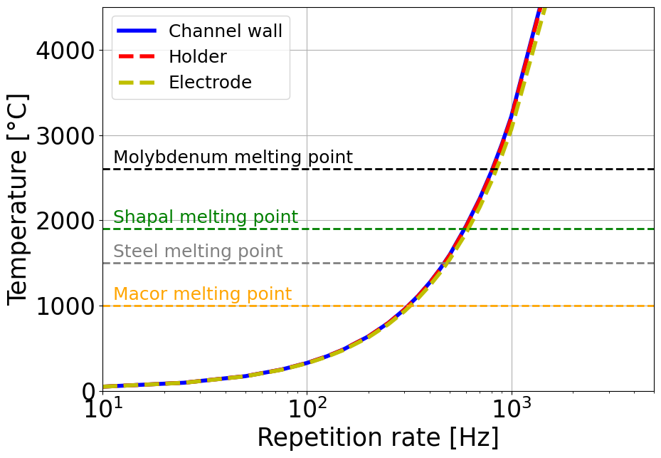}
        \caption{}
        \label{Tvsf}
    \end{subfigure}
    \caption{(a) 3D view of the steady-state surface temperature distribution during operation at 300 Hz. The plasma source is modeled with COMSOL Multiphysics 6.1 \cite{COMSOL}. (b) Equilibrium temperature inside the capillary as a function of the operating repetition rate. The temperature on the channel wall (solid blue line), within the holder (dashed red line) and the electrode (dashed yellow line) are reported together with the melting temperatures of the adopted materials (horizontal dashed lines). %(Left) Equilibrium temperature inside the capillary as a function of the operating repetition rate. (Right) 3D view of the capillary surface temperature during operation at 300 Hz.
    }
    \label{Simu}
\end{figure}

Heat transfer simulations are performed considering the Shapal-Macor capillary shown in Fig.~\ref{Capillary pictures}, subject to a constant heat transfer rate of 1-100 W flowing through the capillary channel walls, thus reproducing the average power deposited by HV pulses in the range 10 Hz - 1 kHz.
As shown in Fig.~\ref{3D view}, the capillary geometry also includes HV cables and the gas injection pipe, conducting the thermal load from the capillary to the HV pulser and the vacuum chamber, which in turn are considered as heat sinks.
In addition, the external surfaces of all the components are thermally insulated, so as to replicate the experimental conditions inside the vacuum chamber.
Due to the heat removal from the source to the heat sink, a thermal steady-state is reached within the capillary.
For instance, Fig.~\ref{3D view} displays the equilibrium temperature distribution within the whole source, with a maximum temperature of 840°C inside the capillary, reached after three hours of continuous plasma discharge operation at 300 Hz.
% as shown in Fig.~\ref{No cooling}.
Moreover, Fig.~\ref{Tvsf} reports the equilibrium temperature as a function of the repetition rate. %that a steady state condition is reached within few hours at a given temperature, depending on the operating repetition rate.

% \begin{figure}[h]
% \centering
% \includegraphics[width=0.99\linewidth]{Images/T plot and 3D image.png}
% \caption{(Left) Equilibrium temperature inside the capillary as a function of the operating repetition rate. (Right) 3D view of the capillary surface temperature during operation at 300 Hz.
% }
% \label{No cooling}
% \end{figure}

In particular, in the range of 10-150 Hz, assessed during the experimental testing, the equilibrium temperature is kept below the melting temperature of Macor and Shapal, thus preventing any damage in the capillary.
For higher repetition rate operation up to the kHz range, the dissipated energy per pulse can be reduced by tuning the discharge duration and peak current, thus keeping the capillary temperature under control.
In conclusion, heat transfer simulations are in good agreement with experimental results, confirming the longevity of the Shapal capillary in high repetition rate operation up to 150 Hz. In addition, simulations provide further insight into the operative limit of the source, which can be extended to the kHz range by properly tuning the experimental settings, such as the discharge current intensity and duration.
In this regard, considering the operative range of 100-400 Hz foreseen by the EuPRAXIA@SPARC\_LAB scientific case, the proposed design of plasma discharge capillaries made in Shapal and Macor represents a reliable solution in terms of longevity and cost-effectiveness.

\section*{Conclusions}

A novel ceramic-based plasma discharge capillary has been tested at 10-150 Hz to assess the reliability of machinable, cost-effective ceramic materials, such as Shapal and Macor, for long-term high repetition rate operation.
Experimental results showed that both plasma properties and the capillary channel profile are preserved during the experimental campaign, proving the ability of the device to dissipate the thermal load produced by the high voltage plasma discharge.
%On the other hand, high repetition rate tests also highlighted that the capillary holder and the electrodes require upgraded designs and suitable materials as well as the capillary itself.
%In this regard the option of a fully ceramic-based device, using Macor (or even Shapal) for the capillary holder, and a composed design for the electrodes, with an inner molybdenum ring and an outer steel plate, proved to be suitable solutions for high repetition rate operation.
Heat conduction simulations confirmed that no excessive overheating takes place during operation up to 150 Hz and the capillary temperature is kept below critical values. In addition, according to the numerical analysis, long-term operation in the kHz range can be achieved by reducing the energy delivered by a single plasma discharge, acting on the discharge duration and current. In particular, experimental and numerical results demonstrate the suitability of Shapal-Macor capillaries to operate in the range 100-400 Hz foreseen for the EuPRAXIA@SPARC\_LAB project.

\section*{Data availability}

The data that support the findings of this study are available from the
corresponding author on reasonable request.

\bibliography{main}

\section*{Acknowledgements}

This work has received funding from the European Union’s Horizon Europe research and innovation program under Grant Agreement No.101079773 (EuPRAXIA Preparatory Phase) and the INFN with the Grant No.GRANT73/PLADIP.
We thank Luca Ioannucci and Gianluca Luminati for the technical support in the installation of the water cooling system for the vacuum pumps and Thomas De Nardis and Gianluca Grilli for the realization of the electrical discharge pulser circuit.

\section*{Author contributions}

A.B. and L.C. proposed the experiment. A.B., L.C. and V.L. designed the ceramic capillary and metal electrodes and installed the vacuum system. D.P. developed the high voltage pulser circuit and the air cooling system. A.B., L.C., R.D. and M.P. carried out the experimental campaign and the plasma source characterization. L.P. and M.R. performed the microscopic analysis of the capillary. L.C. performed heat transfer analysis and numerical simulations. M.F. was involved in the experiment, discussed the results, and reviewed the manuscript.

\section*{Competing interests}

The authors declare no competing interests.

\end{document}